\documentclass[12pt]{article}
\usepackage{amsmath,amssymb,amsthm}

\newcommand\eq[1] {(\ref{#1})}

\newcommand\labfig[1] {\label{fig:#1}}

\newcommand{\beqa}{\begin{eqnarray}}
\newcommand{\eeqa}[1]{\label{#1}\end{eqnarray}}
\newcommand{\beq}{\begin{equation}}
\newcommand{\eeq}[1]{\label{#1}\end{equation}}

\newcommand{\Go}{\omega}

\def\Bf{{\bf f}}

\def\Bn{{\bf n}}

\def\Bu{{\bf u}}

\def\Bx{{\bf x}}

\def\BC{{\bf C}}

\def\BW{{\bf W}}

\def \ba {\begin{array}}
\def \ea {\end{array}}

\def \refe #1.{(\ref{#1})}
\def \reff #1.{figure~\ref{#1}}
\def \refs #1.{section~\ref{#1}}
\def \refss #1.{subsection~\ref{#1}}
\def \refD #1.{Definition~\ref{#1}}
\def \refT #1.{Theorem~\ref{#1}}
\def \refL #1.{Lemma~\ref{#1}}
\def \refC #1.{Corollary~\ref{#1}}
\def \refP #1.{Proposition~\ref{#1}}
\def \refR #1.{Remark~\ref{#1}}
\def \refE #1.{Example~\ref{#1}}
\def \refN #1.{Notation~\ref{#1}}

\usepackage{graphics}
\usepackage{amssymb}
\begin{document}
\vspace{-1in}
\title{A metamaterial having a frequency dependent elasticity tensor and a zero effective mass density}
\author{Graeme Milton\\
\small{Department of Mathematics, University of Utah, Salt Lake City UT 84112, USA}\\ \\
Pierre Seppecher \\
\small{Institut de Math\'ematiques de Toulon}\\
\small{Universit\'e de Toulon et du Var, BP 132-83957 La Garde Cedex, France}}
\date{}
\maketitle
\begin{abstract}
Within the context of linear elasticity we show that a two-terminal network of springs and masses,
can respond exactly the same as a normal spring, but with a frequency dependent spring constant. As 
a consequence a network of such springs can have a frequency dependent effective elasticity tensor but
zero effective mass density. The internal masses influence the elasticity tensor, but do not contribute
to the effective mass density at any frequency.
\end{abstract}
\vskip2mm


\noindent 
\vskip2mm
\section*{Introduction}
       The mass density term which enters the macroscopic wave equation of linear continuum elastodynamics need not be the same
as the mass density calculated by local volume averaging the interior masses. This was observed by Berryman \cite{Berryman:1980:LWPI} in the context 
of an approximation scheme for suspensions of particles in an inviscid fluid and later found to be more generally true for composites constructed from 
components having a high contrast in stiffness and density. The reason is simply that interior masses do not necessarily
locally move together in lock step motion with each other, even when the wavelength is long, and therefore the locally averaged momentum is not simply a product of the 
locally averaged mass times the locally averaged velocity \cite{Sheng:2007:DMD}. At a given frequency, the effective mass density entering the wave equation
can be anisotropic, negative, or even complex \cite{Schoenberg:1983:PPS,Bouchitte:2004:HNR,Liu:2005:AMP,Avila:2005:BPI,Milton:2006:CEM,Sheng:2007:DMD,Milton:2007:MNS}. 
In general, the vibrations of interior masses cause the effective elasticity tensor and the effective mass density to depend on frequency.
This raises the interesting question, which will be explored here, as to whether internal masses
can influence the effective elasticity tensor, and cause it to be frequency dependent, but not contribute to the overall effective mass density at any frequency?
We will see the answer for mass-spring networks is yes, theoretically it can, provided we are strictly working within the framework of linear elasticity and ignore gravity and non-linear instabilities
such as buckling. 
  
\section*{Dispersive Normal Springs}
If we have a normal spring with spring constant $k$ and terminals at the points $\Bx_1$ and $\Bx_2$
then the response of the spring, relating the two forces $\Bf_1$ and $\Bf_2$ applied to the spring at the 
terminals to the displacements $\Bu_1$ and $\Bu_2$ there, takes the form
\beq \begin{bmatrix}\Bf_1 \\  \Bf_2\end{bmatrix}=k\Bn\cdot(\Bu_1-\Bu_2)\begin{bmatrix}\Bn \\  -\Bn\end{bmatrix},
\quad{\rm where}~\Bn=\frac{\Bx_1-\Bx_2}{|\Bx_1-\Bx_2|}.
\eeq{2.1}
We will define a dispersive normal spring
to be any two terminal network of springs and masses such that the relation between 
the forces and displacements at the terminals at any
frequency takes the form \eq{2.1}, but with $k$ being dependent on frequency. In particular, the 
relation \eq{2.1} implies that the forces $\Bf_1$ and $\Bf_2$  are equal in magnitude and opposite
in direction, and directed parallel to $\Bx_1-\Bx_2$.

An essentially explicit scheme for constructing dispersive normal
springs is a corollary of Theorem 4 of Guevara Vasquez, Milton and
Onofrei \cite{Vasquez:2011:CCS}. That theorem covers the much more general case
of characterizing all possible time-dependent responses of multiterminal mass-spring networks,
and generalized earlier work of Camar-Eddine and Seppecher \cite{Camar:2003:DCS} on characterizing
the response of multiterminal spring networks.  Our 
objective here is to present a simpler construction where the mechanism
responsible for the behavior of the network is easy to grasp.

To construct a dispersive normal spring,
consider first the four terminal spring network of figure 1, consisting of a square 
diamond of identical springs with four other identical springs as legs, which 
extend directly outwards from the four vertices of the diamond to the terminals.
This network is a rank one network: it only supports one loading (and all multiples of this loading). To see this
suppose the spring linking terminal 1 to the diamond is under tension $T$.
Then, by balance of forces, the four springs in the diamond must be under tension
$T/\sqrt{2}$ and the other legs joining the diamond to the terminal edges 
must be under tension $T$. Thus forces at all the four terminal nodes are determined if we
know the force at one terminal node : the response of this network, relating
the four forces $\Bf_1$, $\Bf_2$, $\Bf_3$ and $\Bf_4$ at the terminals to the displacements $\Bu_1$, $\Bu_2$, $\Bu_3$ and $\Bu_4$ 
there necessarily takes the form $(\Bf_1, \Bf_2,  \Bf_3,  \Bf_4 )=(T\Bn,  -T\Bn,  T\Bn_\perp, -T\Bn_\perp)$ 
where the scalar tension $T$ depends linearly on $\Bu_1$, $\Bu_2$, $\Bu_3$ and $\Bu_4$ (here $\Bn$ denotes 
the unit vector pointing from  $\Bx_2$ to $\Bx_1$ and $\Bn_\perp$ is the orthogonal unit vector pointing
from $\Bx_4$ to $\Bx_3$). Thus the response matrix $\BW$ which relates the forces at the terminals to the 
displacements there, via the linear relation,
\beq \begin{bmatrix}\Bf_1 \\  \Bf_2 \\  \Bf_3 \\  \Bf_4\end{bmatrix}=\BW
\begin{bmatrix}\Bu_1 \\  \Bu_2 \\  \Bu_3 \\  \Bu_4\end{bmatrix},
\eeq{2.3}
is a rank-one matrix. Since (no matter what the network), 
this response matrix is a positive semidefinite symmetric matrix, we deduce that
\beq \begin{bmatrix}\Bf_1 \\  \Bf_2 \\  \Bf_3 \\  \Bf_4\end{bmatrix}= 
g\begin{bmatrix}\Bn \\  -\Bn \\  \Bn_\perp \\  -\Bn_\perp\end{bmatrix}
\begin{bmatrix}\Bn^T & -\Bn^T & \Bn_\perp^T & -\Bn_\perp^T\end{bmatrix}
\begin{bmatrix}\Bu_1 \\  \Bu_2 \\  \Bu_3 \\  \Bu_4\end{bmatrix}=
g(\Bn\cdot(\Bu_1-\Bu_2)+\Bn_\perp\cdot(\Bu_3-\Bu_4))\begin{bmatrix}\Bn \\  -\Bn \\  \Bn_\perp \\  -\Bn_\perp\end{bmatrix},
\eeq{2.3a}
where the constant $g>0$ scales in proportion to the stiffness of the springs in the
network. Now, let us place a mass $m$ at the terminals 3 and 4, and make them interior
nodes, and consider the response at frequency $\Go$ of the resulting two terminal network.
 The only forces at nodes 3 and
4 are the inertial forces $\Bf_3=m\Go^2\Bu_3$ and $\Bf_4=m\Go^2\Bu_4$. Substituting
this in \eq{2.3a} and eliminating $\Bu_3$ and $\Bu_4$ from the resulting equations,
gives a relation between $(\Bf_1,\Bf_2)$ and $(\Bu_1,\Bu_2)$ exactly of the form
\eq{2.1} with a spring stiffness
\beq k(\Go)=\frac{2gm\Go^2}{m\Go^2-2g}. \eeq{2.4}
Thus this dispersive normal spring behaves exactly like a normal spring but with
a frequency dependent spring stiffness, which will be negative for frequencies
below $\Go_0=\sqrt{2g/m}$. The masses in the
dispersive normal spring do not move, to first order in the displacement, 
when the two terminal nodes undergo a rigid body motion. They only move when the
spring is extended or compressed.

The dispersive normal spring constructed here is floppy. For example, 
within the framework of linear elasticity, the interior diamond can be
infinitesimally rotated with no change in the 
forces $\Bf_1$, $\Bf_2$, $\Bf_3$ and $\Bf_4$ and 
displacements $\Bu_1$, $\Bu_2$, $\Bu_3$ and $\Bu_4$ at the terminals. Strictly speaking one should
perturb the network by adding a scaffolding of additional springs with very small spring constants to uniquely
determine the interior displacements, but we refrain from doing so as to avoid complications. (Since then $\Bf_1$ will not be exactly
equal to $-\Bf_2$.) The undetermined
displacements do not effect the overall response of the dispersive normal spring. However
associated with this degree of
freedom is a buckling mode: when the springs in the network are under compression rather than
tension one can expect, within the framework of non-linear elasticity, that the interior diamond
will rotate, one way or the other, to relieve this compression. We ignore this since we are
working only within the framework of linear elasticity.

We will now forget about the internal structure of the dispersive normal spring and treat it as a single object.

\begin{figure}
\vspace{2in}
\hspace{1.0in}
{\resizebox{2.0in}{1.0in}
{\includegraphics[0in,0in][8in,4in]{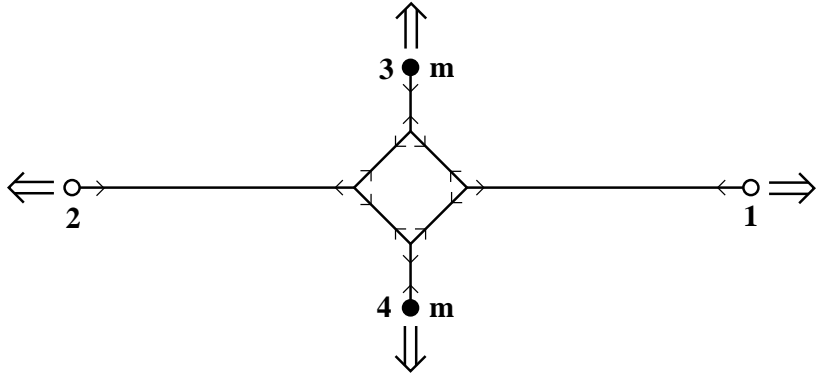}}}
\vspace{0.1in}
\caption{Sketch of the dispersive normal spring. The open circles represent
terminal nodes, and the closed circles could be either terminal nodes or interior nodes with masses
attached. The straight lines represent springs. The large arrows represent external or inertial
forces acting on the nodes at one instant in time. The two small arrows on each spring 
give the direction of the force which the spring exerts on the node nearest to the arrow.}
\labfig{1}
\end{figure}

\section*{A material having frequency dependent elasticity tensors and zero effective
density}
Suppose we have have a periodic network (such as a triangular network) 
of identical normal springs with spring stiffness $k$. Let the nodes of this
network be called ``primary nodes''. The effective elasticity
tensor of the network $\BC$ will be proportional to $k$ and we can write $\BC=k\BC^0$.
Now if we replace each spring in the network by a dispersive normal spring
with the same spring stiffness $k(\Go)$ given by \eq{2.4}, then the resulting material
will have a frequency dependent effective elasticity tensor
\beq \BC=k(\Go)\BC^0=2gm\Go^2\BC^0/(m\Go^2-2g),
\eeq{3.3}
which is negative definite for frequencies
below $\Go_0=\sqrt{2g/m}$.

At the same time, within the framework of linear elasticity, 
the effective density of this network will be zero! By this we mean
that at any fixed frequency, and to first order in the displacements, that the
macroscopic response of the network will be the same as that of a
periodic network of identical normal springs, without masses at the nodes.
The internal masses 
cause the effective elasticity tensor to depend on frequency (and cause it to 
be negative definite for $\Go<\Go_0$) but do not contribute to the effective
density, since the internal masses do not move (to first order in the displacements)
when the primary node lattice is translated.    
Of course, as opposed to normal linear elasticity which can be useful even if there are large
displacements such as rotations, the linear elasticity approximation here
will only be valid for displacements which are small compared to the
length of the springs in the network, if at all. 

Some care needs to be applied in this notion of effective elasticity tensors.
If we take a large sample of the periodic network of dispersive normal springs
then we need to ensure that any cut dispersive normal spring at the
boundary of the sample is removed and that there are no interior nodes
of the dispersive normal springs in contact with the surface loadings applied
to the boundary of the sample.
If there are body forces (perhaps due to
electric field gradients and polarizable nodes) then we need to make sure
these only act on the primary nodes common to the original network of normal springs and
not on the internal nodes and masses of each individual dispersive normal spring.

\section*{Acknowledgements}
GWM is grateful for support from the University of Toulon-Var and from the
National Science Foundation through grant DMS-0707978.

\bibliographystyle{pss}
\bibliography{/u/ma/milton/tcbook,/u/ma/milton/newref}

\end{document}